\documentclass[11p]{article}
\usepackage{amsthm,amsmath,amssymb,hyperref,multirow,graphicx,color}
\newcommand{\Z}{\mathbb{Z}}

\begin{document}
\title{Algebraic and Topological Indices of Molecular Pathway Networks in Human Cancers}
\author{Peter Hinow\thanks{Department of Mathematical Sciences, University of Wisconsin - Milwaukee, P.O. Box 413, Milwaukee, WI 53201, USA;  \texttt{hinow@uwm.edu}} \and 
Edward A.~Rietman\thanks{Newman-Lakka Institute, Tufts University School of Medicine, Boston,  MA 02111, USA;  \texttt{erietman@gmail.com}} \and 
Jack A.~Tuszy\'nski\thanks{Cross Cancer Institute and Department of Physics, University of Alberta, Edmonton, T6G 2E1, Canada;  \texttt{jack.tuszynski@gmail.com}}}
\maketitle

\begin{abstract} 
\begin{sloppypar}
Protein-protein interaction networks associated with diseases have gained prominence as an area of research.
We investigate algebraic and topological indices for protein-protein interaction networks of 11 human cancers derived from the Kyoto Encyclopedia of Genes and Genomes (KEGG) database. We find a strong correlation 
between relative automorphism group sizes and topological network complexities on the one hand and  five year survival probabilities on the other hand. Moreover, we identify several protein families (e.g.~PIK, ITG, AKT families)  that are repeated motifs in many of the cancer pathways. Interestingly, these sources of symmetry are often central rather than peripheral. Our results can aide in identification of promising targets for anti-cancer drugs. Beyond that, we provide a unifying framework to study protein-protein interaction networks of families of related diseases (e.g.~neurodegenerative diseases, viral diseases, substance abuse disorders).
\end{sloppypar}
\end{abstract}


\section*{Introduction}
Biological networks have been an active area of research for some years, see e.g.~\cite{Junker_2008,Koonin_2006,Maayan_2012} and the references therein. In earlier work \cite{Breitkreutz_2012}  we reported that molecular signaling network complexity is correlated with cancer patient survival. In that work we reported  a statistical mechanics measure of network complexity. Here we focus on the relative sizes of automorphism  groups and the dimensions of the cycle spaces (cyclomatic numbers).

Complex real-world networks contain feedback loops to enable the network ``communication'' to continue in the face of node failure \cite{Albert_2000}. In the case of  protein-protein interaction (PPI) networks this means that inhibition of a specific node may or may not have any effect. It is well known that targeting hub nodes in networks often causes the network to break up into multiple components and this could be lethal, because many protein hubs in PPIs for cancer are also important proteins in metabolic networks. As we argued in \cite{Breitkreutz_2012}, targeting nodes with high-betweenness has higher potential for improved cancer treatment. Selective targeting of nodes in a PPI for cancer treatment is fraught with difficulties.

In this letter  we apply two more algebraic and topological indices to study cancer PPI networks and show correlation with 5 year patient survival. We identify several repeated motifs of proteins that are ``interchangeable'' in a sense to be specified below.  In the long run we anticipate that the methods described here will aid identification  of potential drug targets.

\section*{Results and Discussion}
A {\it network} is an undirected graph $G=(V,E)$ with vertex set $V$  and edge set $E$. The vertices are proteins and two vertices are connected by an edge if there is a known interaction of the two partners, either by direct binding or by enzymatic catalysis. Beyond cancer pathways, the Kyoto Encyclopedia of Genes and Genomes (KEGG) database also contains pathways related to immune diseases (e.g.~asthma), neurodegenerative diseases (Alzheimer's disease, Parkinson's disease), substance dependence, cardiovascular diseases, viral diseases and many others  \cite{Kanehisa_2004}. The KEGG networks are assembled from  the literature by searches for experimental confirmation of the relevant interactions. Each interaction is always confirmed by two or more different experimental techniques such as pull-down mass spectrometry, yeast two-hybrid and various biochemical tests. Naturally, networks constructed from experimental results are likely to contain errors, which are however impossible to quantify.

An {\it automorphism} is a permutation  $\phi:V\to V$ that preserves the adjacency relation, that is,
\begin{equation*}
(u,v)\in E \:\: \Leftrightarrow (\phi(u),\phi(v))\in E. 
\end{equation*}
With the operation of composition, the automorphisms form a group $Aut(G)$.  The relation on the set of vertices
\begin{equation*}
u\sim v \:\: \Leftrightarrow \textrm{ there exists a } \phi\in Aut(G)\textrm{ such that } v=\phi(u)
\end{equation*}
is an equivalence relation and its equivalence classes are called the (group) orbits.
 
MacArthur {\it et al.}~\cite{MacArthur_2008} list 20 examples of real world networks and their rich symmetry groups. This is in contrast to  large random graphs, such as graphs from the Erd\H{o}s-R\'enyi model $\mathcal{G}(n,p)$. Here $n$ is the number of vertices. Edges are independently present with probability $0<p<1$. Such graphs have only the trivial automorphism, with probability approaching one, in the limit $n\to\infty$ \cite[Chapter IX]{Bollobas_2001}. The difference is not   surprising if one realizes that real networks display a modular structure, with vertices organized in communities tightly connected internally and loosely connected to each other \cite{Garlaschelli_2010}. This results in  the presence of symmetric subgraphs such as trees and complete cliques. 

\begin{figure}[h!]
\centering
\includegraphics[width=120mm]{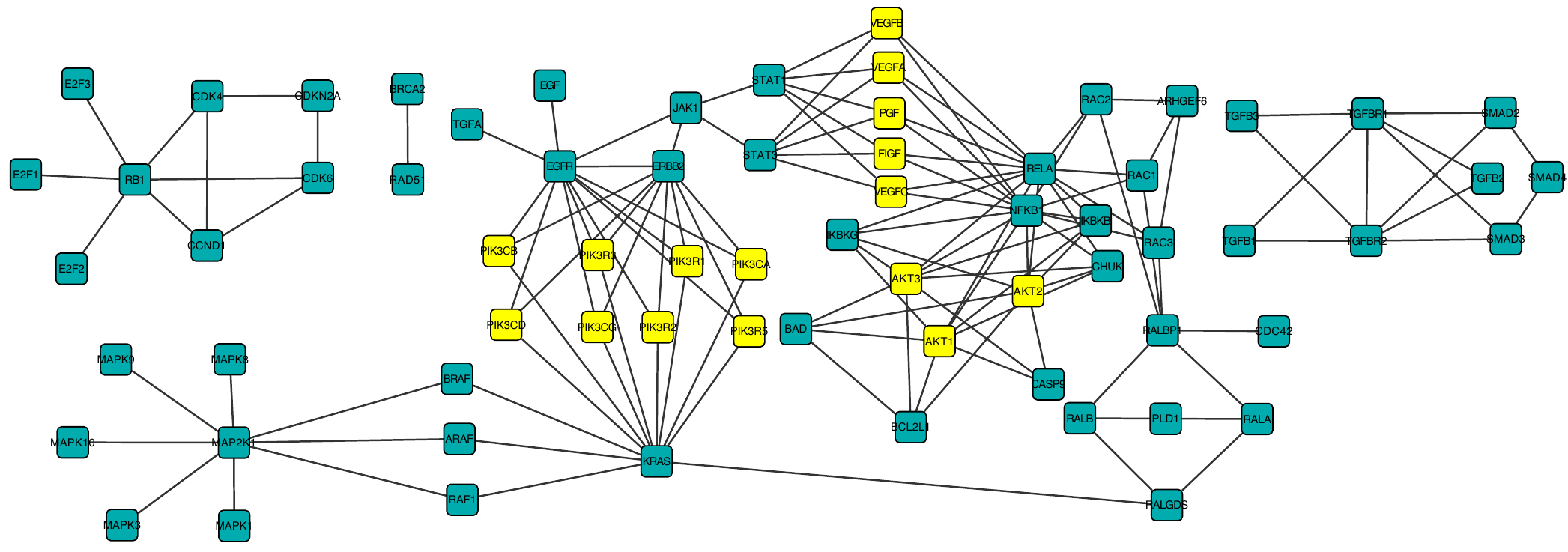}
\caption{\textit{The protein-protein interaction network of pancreatic cancer.} The  network was retrieved from the KEGG database \cite{Kanehisa_2004}  and its automorphism group determined with  {\tt saucy} \cite{Katebi_2012}, namely $Aut(G) = S_2^9\times S_3^6\times S_5^2 \times S_8$. 
Highlighted in yellow are three central orbits of nodes of  degrees 3, 4 and 8, respectively. Two of these are the PI3K and the AKT families, respectively.}\label{pancreatic}
\end{figure}

Figure \ref{pancreatic} shows as an example the protein-protein interaction network of pancreatic cancer as retrieved from the KEGG database. We find that the automorphism group of this network is the direct product of symmetric groups
\begin{equation*}
Aut(G) = S_2^9\times S_3^6\times S_5^2 \times S_8,
\end{equation*}
see Table \ref{AllCancers} for a complete list of automorphism groups. Remarkably, symmetries do not only  arise due to tree subgraphs at the ``ends'' of the network, but also due to central nodes of high degree (highlighted in yellow in Figure \ref{pancreatic}). Thus any flow of information that passes through one node in such an orbit equivalence class may pass through any other node in the same equivalence class. The presence of such modular patterns indicates a high level of redundancy  which confers robustness to the associated biological system (tumor cells).  We suggest that to interrupt the flow through such a network most efficiently, the nodes adjacent to large central orbits are the best to be targeted for example by pharmacological agents that inhibit a specific protein-protein interaction pair. Similar suggestions have been made in \cite{Chandra_2013,Csermely_2013,Winterbach_2013}. The use of automorphism groups has, to the best of our knowledge, not yet been proposed.

Automorphism groups are often used to measure the complexity of a network \cite{Xiao_2008}. In order to make automorphism group sizes of  graphs  with $n=|V|$ vertices comparable, we follow the suggestion in  \cite{Xiao_2008} and compute the ratio 
\begin{equation*}
\beta_G=\left(\frac{|Aut(G)|}{n!}\right)^\frac{1}{n}.
\end{equation*}
This relates the size of $Aut(G)$ to the size of the automorphism group $S_n$ of the complete graph on $n$ vertices.

\begin{figure}[h!]
\centering
\includegraphics[width=70mm,height=60mm]{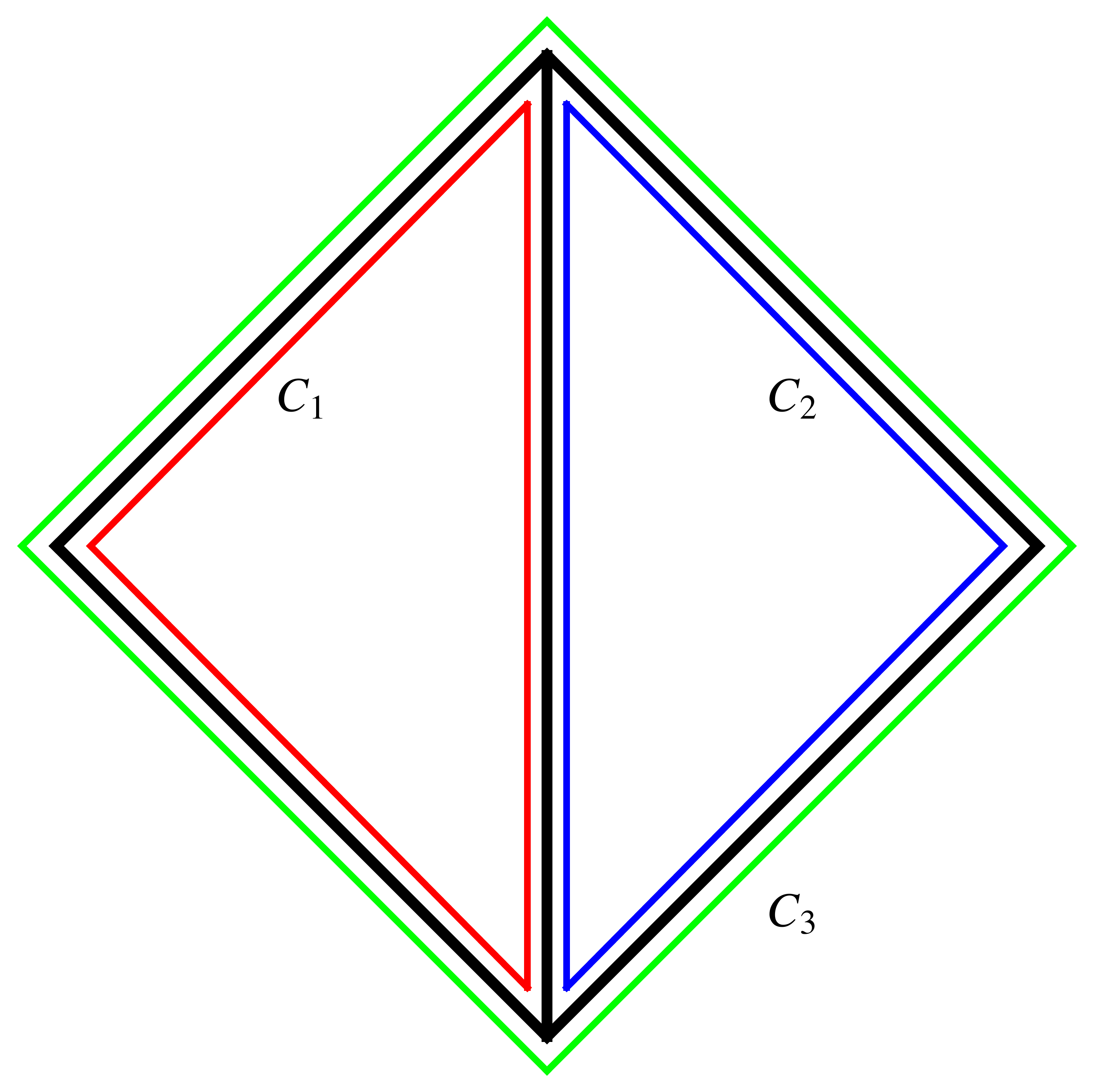}
\caption{\textit{The cycle space of a simple graph with four vertices.} The graph $G$ is shown in black and three cycles $C_1,\,C_2$ and $C_3$ are marked in red, blue and green, respectively. Any of the sets $\{C_1,C_2\},\,\{C_1,C_3\}$ and  $\{C_2,C_3\}$ is a basis of the cycle space $\mathcal{C}(G)=\{C_1,C_2 ,C_3\}$.}\label{cyc_sp}
\end{figure}

A second graph invariant is of a more topological nature. A {\it cycle} is a sequence of adjacent vertices that starts and ends at the same vertex. 
The set of all cycles $\mathcal{C}(G)$ can be made a vector space over the field $\Z_2$ by taking the symmetric difference 
\begin{equation*}
C_1\Delta C_2 = (C_1\cup C_2) \setminus (C_1 \cap C_2)
\end{equation*}
as addition, the identity as negation, and the empty cycle as zero. The dimension of this vector space is called the    {\it cyclomatic number} $\mu_G$,  or the {\it circuit rank}. Loosely speaking, it is a count of the ``independent'' loops, see Figure \ref{cyc_sp}.  It is shown in \cite{Berger_2004,Kavitha_2009} that for a graph with $n$ vertices, $m$ edges and $c$ connected components, $\mu_G$ is given by 
\begin{equation*}
\mu_G=m-n+c.
\end{equation*}

\begin{figure}[h!]
\centering
\includegraphics[width=120mm]{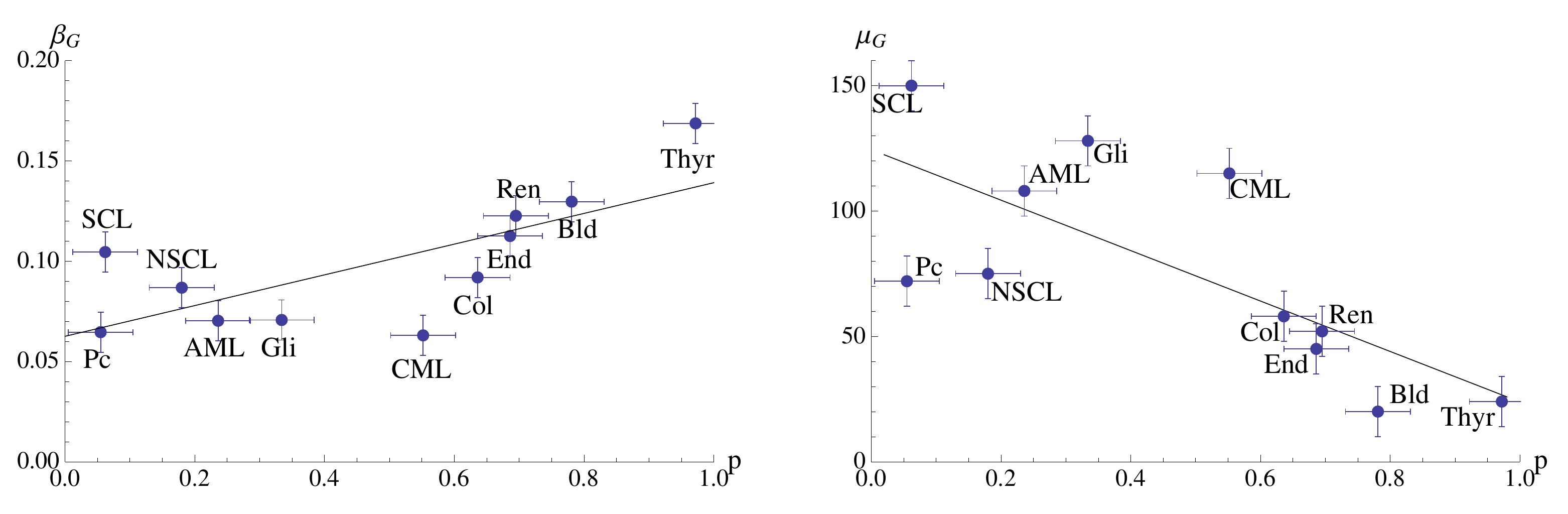}
\caption{\textit{Automorphism group sizes  and cyclomatic numbers.}
A plot of relative automorphism group size $\beta_G$  (left) and cyclomatic number $\mu_G$ (right) against five year survival probability $p$ for 11 types of cancer. In both cases we have  the coefficients of determination  $R^2 = 0.52$ and  $p = 0.011$. The cancer types are {\it AML}: acute myelogeneous leukemia, {\it Bld}: bladder cancer,   {\it CML}: chronic myelogeneous leukemia,  {\it Col}: colorectal cancer,  {\it End}: endometrial cancer,  {\it Gli}: Glioma,  {\it NSCL}: non-small cell lung cancer,  {\it Pc}: pancreatic cancer,  {\it Ren}: renal cancer,  {\it SCL}: small cell lung cancer,  {\it Thyr}: thyroid cancer.  The width of the horizontal error bars is 0.1, the width of the vertical error bars is 0.02 in the left panel and 20 in the right panel. These errors are estimated as the actual error is unknown.}\label{correlation}
\end{figure}

\begin{figure}[h!]
\centering
\includegraphics[width=70mm]{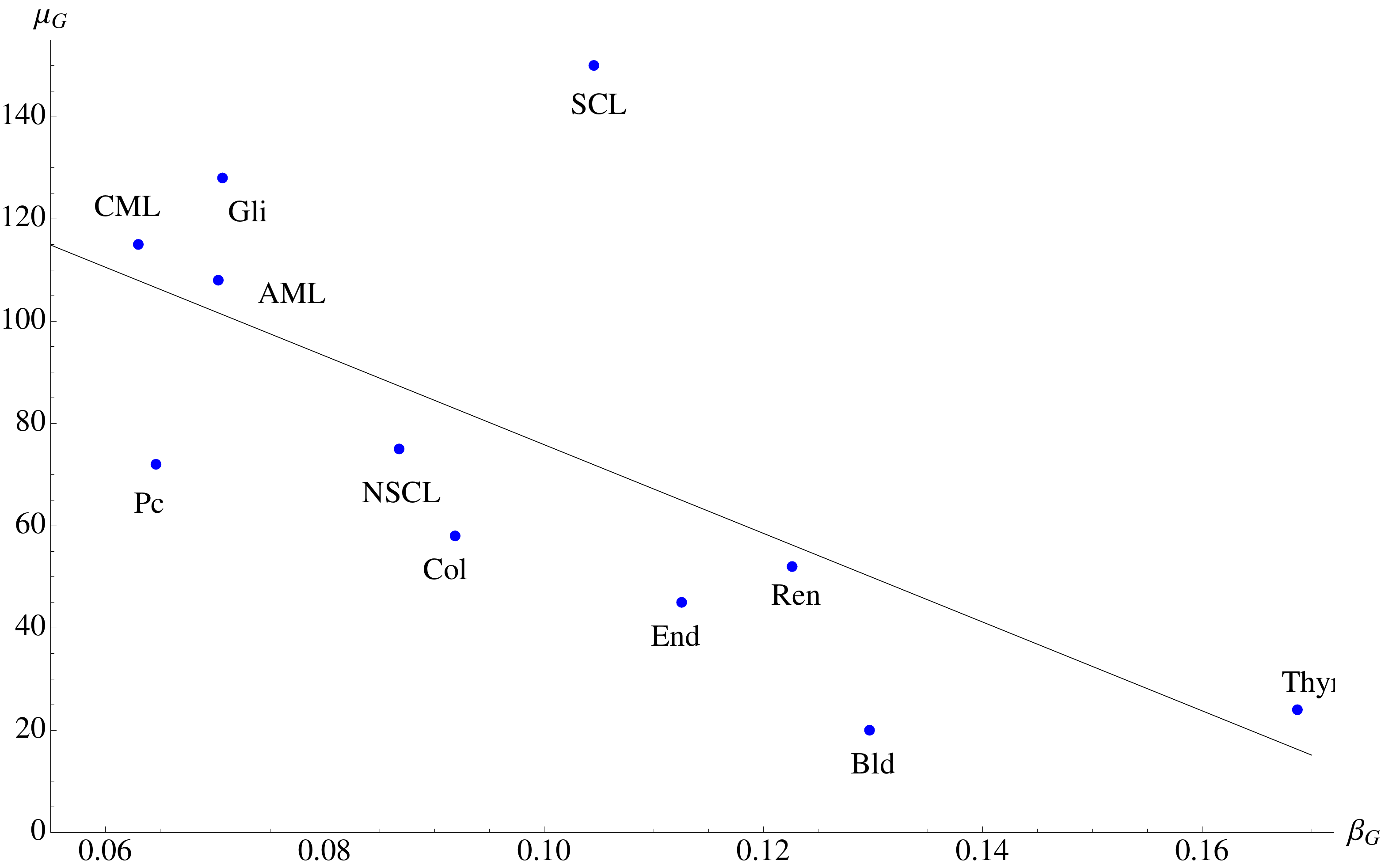}
\includegraphics[width=40mm,height=47mm]{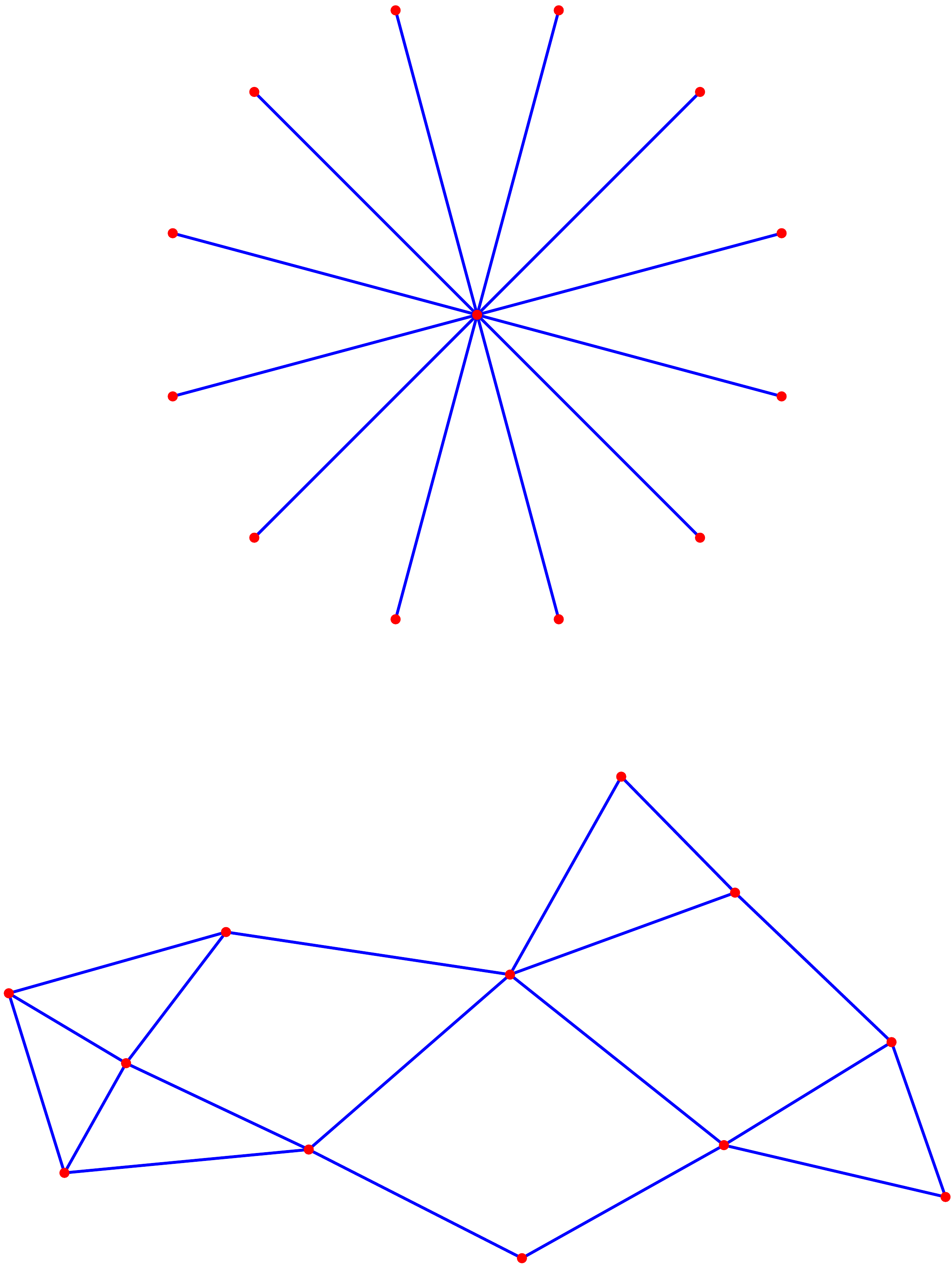}
\caption{\textit{Mutual relation between $\beta_G$ and $\mu_G$.} (Left) 
A plot of  cyclomatic number $\mu_G$ against relative automorphism group size $\beta_G$ for the 11  protein-protein interaction networks of cancers. (Right) In general, these quantities are not related. Shown are a highly symmetric graph with trivial cycle space and an asymmetric graph with large cycle space.}\label{no_relation}
\end{figure}

We plot these two indices against the five year survival probability $p$, obtained from the Surveillance, Epidemiology and End Results (SEER) database \cite{SEER_2013} for 11 types of cancer in Figure \ref{correlation}. Interaction networks with larger values of $\beta_G$ or equivalently greater symmetry are associated with better chances of survival. A large value of $\mu_G$ indicates high topological complexity and correlates with decreased chance of survival. We find that  both coefficients of determination are $R^2 = 0.52$ with corresponding $p$-value $p = 0.011$ (the equality is coincidental). There are widespread differences in detection stage, metastasis status, treatment and general health of the patient which are unfortunately not accessible from the SEER database. Nevertheless, given this large amount of natural uncertainty in the data, this indicates a strong correlation of averages. It would be invaluable for future research to classify database entries according to some of the parameters mentioned above.  Since both the automorphism group size  $\beta_G$ and the   cyclomatic number $\mu_G$ are correlated to the  five year survival probability $p$, it is to be expected that these two quantities are correlated to each other, for the  the protein-protein interaction networks of cancers that are the object of our study, see Figure \ref{no_relation}, left panel. However, it is easy to construct examples of graphs that show no correlation between  $\beta_G$ and   $\mu_G$, see Figure \ref{no_relation}, right panel.  

Further study of the automorphism groups reveals  repeated motifs in several interaction networks. The eight proteins from the PIK3C\{A,B,D,G\} and  PIK3R\{1,2,3,5\} family form a single orbit equivalence class in seven of the networks (AML, CML, colorectal, endometrial, pancreatic, renal and SCL cancers) and are split in two orbit equivalence class in two more networks (glioma and NSCL). The three proteins AKT\{1,2,3\} are orbit equivalent in eight networks (CML, colorectal, endometrial, glioma, NSCL, pancreatic, renal and SCL cancers), that is, whenever they appear in the network to begin with. These players have been known for a long time to be of crucial importance to the initiation and progression of cancer, mainly due to the various biological and biochemical assays performed on cancer cells. However, our conclusions stem directly from a group-theoretic analysis of the PPI networks and they are network specific.  Since its initial discovery as a proto-oncogene, the serine/threonine kinase AKT has become a major focus of attention because of its critical regulatory role in diverse cellular processes, including cancer progression and insulin metabolism. The AKT cascade is activated by receptor tyrosine kinases, integrins, B and T cell receptors, cytokine receptors, G-protein-coupled receptors and other stimuli that induce the production of phosphatidylinositol (3,4,5)-triphosphates (PtdIns(3,4,5)P3) by phosphoinositide 3-kinase (PI3K). These lipids serve as plasma membrane docking sites for proteins that harbor pleckstrin-homology (PH) domains, including AKT and its upstream activator PDK1. The tumor suppressor PTEN is recognized as a major inhibitor of AKT and is frequently lost in human tumors. There are three highly related isoforms of AKT (AKT1, AKT2, and AKT3), which represent the major signaling arm of PI3K. For example, germline mutations of AKT have been identified in pathological conditions of cancer and insulin metabolism. AKT regulates cell growth through its effects on the TSC1/TSC2 complex and mTOR pathways, as well as cell cycle and cell proliferation through its direct action on the CDK inhibitors p21 and p27, and its indirect effect on the levels of cyclin D1 and p53. AKT is a major mediator of cell survival through direct inhibition of pro-apoptotic signals such as the pro-apoptotic regulator BAD and the FOXO and Myc family of transcription factors. AKT has been demonstrated to interact with Smad molecules to regulate TGF-$\beta$ signaling. These findings make AKT an important therapeutic target for the treatment of cancer.

Interestingly, the network of small cell lung cancer contains an enormous orbit of 18 equivalent nodes of degree six. This orbit consists of laminines, collagens and a fibronectin that  are major proteins in the basal lamina. All nodes are connected to six members of the integrin family of transmembrane receptors.

\section*{Conclusion}

We have shown that the relative size of the automorphism groups and the cyclomatic numbers for cancer pathway networks from the  KEGG  database are both correlated with five-year survival of cancer patients. Determination of the specific reasons for these great discrepancies in survival rates remains a topic for future research. Interestingly, cancers with more symmetric interaction networks are associated with better survival rates. This may be due to a greater robustness to failure, which, somewhat counterintuitive, is a positive feature in this context. 

We suggest that selective removal of nodes from the network (clinically equivalent to protein inhibition) and reinterpolation on the linear curves helps to identify  potential drug targets. This indicates that complexity of a biochemical network involved in a deregulated cell cycle as exemplified by cancer cells is of crucial importance to its robustness. This is manifested by various redundancies in the PPI network that make the search for a therapeutic ``silver bullet'' an impossible task. We suggest that selective removal of nodes from the network (clinically equivalent to protein inhibition) and reinterpolation on the linear curves helps to identify potential drug targets. We have shown that PI3K and AKT families of proteins appear to be the most suitable targets for pharmacological inhibition in the most number of cancer types studied. It is encouraging that there are several AKT pathway inhibitors in clinical development, e.g. perifosine (KRX-0401, Aeterna Zentaris/Keryx), MK-2206 (Merck), and GSK-2141795 (Glaxo-SmithKline) \cite{Alexander_2011}. Similarly, Bayer, GlaxoSmithKline (GSK), Novartis, Merck \& Co., Roche and Sanofi are just a few of the companies that have placed great importance   on the development of a spectrum of agents targeting the PI3K pathway. Drug candidates including pan-PI3K inhibitors, PI3K isoform-specific inhibitors, AKT inhibitors and mTOR inhibitors are currently tested alone and in combinations in an array of cancer indications \cite{Holmes_2011}. While the motivation for this focus has been stated as: “The pathway is almost invariably on in cancer”, our methodology identifies this pathway as the most crucial using mathematical analysis of the network. Moreover, we are able to identify those types of cancer where the pathway should be the main target and those types where targeting it may not produce the expected clinical outcomes.

\section*{Methods}
The cancer pathways were obtained from the Kyoto Encyclopedia of Genes and Genomes (KEGG) \cite{Kanehisa_2004}  with the help of the open source software packages {\tt KEGGgraph} \cite{Zhang_2009} and {\tt cytoscape} \cite{Shannon_2003}.   The automorphism groups of the networks were  found with  {\tt saucy} \cite{Katebi_2012} and {\tt gap} \cite{GAP_2013} (see Tables \ref{AML}-\ref{thyroid}  for the  complete group lists). Bases of the cycle spaces were found using {\tt python} networkX.

\section*{Acknowledgements}
JAT acknowledges funding support for this research from NSERC (Canada), Canadian Breast Cancer Foundation and the Allard Foundation.
We thank Dr.~Giannoula Klement (Newman-Lakka Institute, Tufts University School of Medicine) for valuable comments.

\section*{Figures}

\section*{Tables }
\begin{table}[h!]
\centering
\begin{tabular}{|l|c|l|c|c|c|c|} \hline
Cancer & $n$ & $Aut(G)$ & $|Aut(G)|$ & $c$ & $\beta_G$ & $\mu_G$ \\ \hline
AML & 62   & $S_2^7\times S_3^4\times S_4 \times S_7^2$  & $1.011\cdot 10^{14}$  & 1  & $7.03\cdot 10^{-2}$ & 108  \\ \hline
Bld & 29   & $S_2^5\times S_3^3\times S_4$  & $1.658\cdot 10^5$  &  3  & $1.29\cdot 10^{-1}$ & 20  \\ \hline
CML & 73   & $S_2^8\times S_3^7\times S_4 \times S_5^2 \times S_8$  & $9.98\cdot 10^{17}$  &  3  & $6.3\cdot 10^{-2}$ & 115  \\ \hline
Col & 49   & $S_2^5\times S_3^4\times S_4^2 \times S_8$  & $9.63\cdot 10^{11}$  &  3   & $9.2\cdot 10^{-2}$ & 58  \\ \hline
End & 44   & $S_2^5\times S_3^4\times S_4\times S_5 \times S_8$  & $ 4.82\cdot 10^{12}$  &  2   & $1.1\cdot 10^{-1}$ & 45  \\ \hline
Gli & 64   & $S_2^8\times S_3^6\times S_4^4\times S_6$  & $ 2.85\cdot 10^{15}$  &  2   & $7.0\cdot 10^{-2}$ & 128  \\ \hline
NSCL & 48   & $S_2^7\times S_3^5\times S_4^3$  & $1.37\cdot 10^{10}$  &  1   & $8.6\cdot 10^{-2}$ & 75  \\ \hline
Pc & 69   & $S_2^9\times S_3^6\times S_5^2\times S_8$  & $1.39\cdot 10^{16}$  &   4   & $6.4\cdot 10^{-2}$ & 72  \\ \hline
Ren & 60   & $S_2^8\times S_3^2\times S_6\times S_8\times S_{18}$  & $1.7\cdot 10^{27}$  &  3   & $1.2\cdot 10^{-1}$ & 52  \\ \hline
SCL & 77   & $S_2^4\times S_3^4\times S_4\times S_6\times S_8\times S_{12}\times S_{18}$  & $4.4\cdot 10^{37}$  &  2   & $1.0\cdot 10^{-1}$ & 150  \\ \hline
Thyr & 28   & $S_2^4\times S_3^2\times S_4\times S_7$  & $7.0\cdot 10^{7}$  &  3   & $1.7\cdot 10^{-1}$ & 24  \\ \hline
\end{tabular}
\caption{Automorphism groups of all cancers. Column $n$ contains the number of vertices, column $c$ contains the number of connected components of the protein-protein interaction network. The abbreviations are the same as in Figure \ref{correlation}.}\label{AllCancers}
\end{table}

\begin{table}[h!]
\begin{tabular}{|l|l|l|l|l|l|l||c|} \hline
\multicolumn{7}{|c||}{Nodes} & Degree \\ \hline
PIK3R5 & PIK3CA & PIK3CB & PIK3CD & PIK3R1 & PIK3R2 & PIK3R3 & 9 \\ \hline
CCND1 & MYC & PPARD & LEF1 & TCF7 & TCF7L2 & TCF7L1 & 1 \\ \hline
KRAS & NRAS & HRAS &   &  &  &  & 12 \\ \hline
RPS6KB1 & RPS6KB2 & EIF4EBP1 &   &  &  &  & 5 \\ \hline
STAT3 & STAT5A & STAT5B &   &  &  &  & 5 \\ \hline
RARA & PML & LOC652346 & LOC652671 &  &  &  & 3 \\ \hline
ARAF & RAF1 & BRAF &   &  &  &  & 1 \\ \hline
KIT & LOC652799 &  &   &  &  &  & 12 \\ \hline
IKBKB & IKBKG &  &   &  &  &  & 5 \\ \hline
MAP2K1 & MAP2K2 &  &   &  &  &  & 5 \\ \hline
SOS1 & SOS2 &  &   &  &  &  & 4 \\ \hline
PIM1 & PIM2 &  &   &  &  &  & 3 \\ \hline
NFKB1 & RELA &  &   &  &  &  & 3 \\ \hline
MAPK1 & MAPK3 &  &   &  &  &  & 2 \\ \hline
\end{tabular}
\caption{Orbit equivalence classes for the AML group.}\label{AML}
\end{table}

\begin{table}[h!]
\begin{tabular}{|l|l|l|l||c|} \hline
\multicolumn{4}{|c||}{Nodes} & Degree \\ \hline
DAPK1 & DAPK3 & DAPK2 & RPS6KA5   & 2 \\ \hline
KRAS & NRAS & HRAS &   & 5 \\ \hline
ARAF & RAF1 & BRAF &    & 5 \\ \hline
E2F1 & E2F2 & E2F3 &  &1 \\ \hline
MAPK1 & MAPK3 &   &  & 6 \\ \hline
MAP2K1 & MAP2K2 &   &  &  5\\ \hline
RASSF1 & FGFR3 &   & &3 \\ \hline
CCND1 & CDK4 &   & &3 \\ \hline
EGFR & ERBB2 &   & &1 \\ \hline
\end{tabular}
\caption{Orbit equivalence classes for the  bladder cancer group.}\label{BlCanc}
\end{table}

\begin{table}[h!]
\begin{tabular}{|l|l|l|l|l|l|l|l||c|} \hline
\multicolumn{8}{|c||}{Nodes} & Degree \\ \hline
PIK3R5 & PIK3CA & PIK3CB & PIK3CD & PIK3CG & PIK3R1 & PIK3R2 & PIK3R3 & 9 \\ \hline
CRK & CRKL & CBLC & CBL & CBLB  &  &  & & 9 \\ \hline
SHC2	 & SHC4 & 	SHC3 & SHC1 & MYC  &  &  &  & 2 \\ \hline
HDAC1 & HDAC2 & CTBP1 & CTBP2 &   &  &  &  & 2 \\ \hline
AKT1 & AKT2 & AKT3 &  &   &  &  &  & 14 \\ \hline
ARAF & RAF1 & BRAF &   &  &  & &  & 5 \\ \hline
KRAS & NRAS & HRAS &   &  &   &  &  & 5 \\ \hline
CHUK & IKBKB & IKBKG &  &   &  &  &  & 4 \\ \hline
CDK4 & CDK6 & CCND1 &  &   &  &  &  & 3 \\ \hline
E2F1 & E2F2 & E2F3 &  &   &  &  &  & 1 \\ \hline
ABL1 & BCR &  & &  &  &  &  & 13 \\ \hline
MAP2K1 & MAP2K2 &  & &  &  &  &  & 5 \\ \hline
SOS1 & SOS2 &  &  & &  &  &  & 4 \\ \hline
MECOM & RUNX1 &  &  & &  &  &  & 4 \\ \hline
STAT5A & STAT5B &   & &  &  &  &  & 3 \\  \hline
MAPK1 & MAPK3 &  & &  &  &  &  & 2 \\ \hline
TGFRB1 & TGFRB2 &  & &  &  &  &  & 2 \\ \hline
NFKB1 & RELA &  &  & &  &  &  & 1 \\ \hline
\end{tabular}
\caption{Orbit equivalence classes for the CML group.}\label{CML}
\end{table}

\begin{table}[h!]
\begin{tabular}{|l|l|l|l|l|l|l|l||c|} \hline
\multicolumn{8}{|c||}{Nodes} & Degree \\ \hline
PIK3R5 & PIK3CA & PIK3CB & PIK3CD & PIK3CG & PIK3R1 & PIK3R2 & PIK3R3 & 4 \\ \hline
RHOA & RAC1 & RAC2 & RAC3 &   &  &  &  & 4 \\ \hline
LEF1 & TCF7 & TCF7L1 &  TCF7L2 &   &  &  &  & 4 \\ \hline
AKT1 & AKT2 & AKT3 &  &   &  &  &  & 8 \\ \hline
MAPK8 & MAPK9 & MAPK10  & &  &  &  &  & 4 \\ \hline
ARAF & RAF1 & BRAF &   &  &  & &  & 2 \\ \hline
SMAD4 & TGFBR1 & TGFBR2 &   &  &   &  &  & 2 \\ \hline
MAPK1 & MAPK3 &  & &  &  &  &  & 4 \\ \hline
JUN & FOS &  &  & &  &  &  & 3 \\ \hline
DCC & CASP3 &  & &  &  &  &  & 3 \\ \hline
SMAD2 & SMAD3 &  &  & &  &  &  & 3 \\ \hline
\end{tabular}
\caption{Orbit equivalence classes for the colorectal cancer group.}\label{ColRec}
\end{table}

\begin{table}[h!]
\begin{tabular}{|l|l|l|l|l|l|l|l||c|} \hline
\multicolumn{8}{|c||}{Nodes} & Degree \\ \hline
PIK3R5 & PIK3CA & PIK3CB & PIK3CD & PIK3CG & PIK3R1 & PIK3R2 & PIK3R3 & 3 \\ \hline
APC2	& APC & AXIN1 & AXIN2 & GSK3B &  &  &  & 1 \\ \hline
LEF1 & TCF7 & TCF7L1 &  TCF7L2 &   &  &  &  & 3 \\ \hline
AKT1 & AKT2 & AKT3 &  &   &  &  &  & 5 \\ \hline
KRAS & NRAS & HRAS &   &  &   &  &  & 5 \\ \hline
ARAF & RAF1 & BRAF &   &  &  & &  & 5 \\ \hline
BAD & CASP9 & FOXO3 &   &  &   &  &  & 3 \\ \hline
PDPK1 & ILK &  &  & &  &  &  & 10 \\ \hline
MAP2K1 & MAP2K2 &  & &  &  &  &  & 5 \\ \hline
MAPK1 & MAPK3 &  & &  &  &  &  & 3 \\ \hline
SOS1 & SOS2 &  &  & &  &  &  & 4 \\ \hline
MYC & CCND1 &  &  & &  &  &  & 4 \\ \hline
DCC & CASP3 &  & &  &  &  &  & 3 \\ \hline
\end{tabular}
\caption{Orbit equivalence classes for the endometrial cancer group.}\label{EndomC}
\end{table}

\begin{table}[h!]
\begin{tabular}{|l|l|l|l|l|l||c|} \hline
\multicolumn{6}{|c||}{Nodes} & Degree \\ \hline
CALML6 & CALML5 & CALM1 & CALM2 & CALM3 & CALML3  & 3 \\ \hline
PIK3CA & PIK3CB & PIK3CD & PIK3CG & & & 9\\ \hline
CAMK2A & CAMK2B & CAMK2D &  CAMK2G &   &  & 8 \\ \hline
PIK3R5 & PIK3R1 & PIK3R2 & PIK3R3 & & & 7\\ \hline
SHC1	& SHC2 & SHC3 & SHC4 &  &  & 5 \\ \hline
KRAS & NRAS & HRAS &      &  &  & 20 \\ \hline
ARAF & RAF1 & BRAF &   &  &  & 5 \\ \hline
PRKCA & PRKCB & PRKCG &  &     &  & 3 \\ \hline
CCND1 & CDK4 & CDK6 & & & &3 \\ \hline
AKT1 & AKT2 & AKT3 &  &     &  & 1 \\ \hline
E2F1 & E2F2 & E2F3 &  & &  &  1 \\ \hline
PDGFRA & PDGFRB &  &  &  &  & 16 \\ \hline
MAP2K1 & MAP2K2 &  &  &  &  & 9 \\ \hline
PLCG1 & PLCG2 &  &  &  &  & 4 \\ \hline
SOS1 & SOS2 &  &  &  &  & 4 \\ \hline
\multicolumn{2}{|l|}{MDM2 $\leftrightarrow$ TP53 (2)} & & \multicolumn{2}{|l|}{CDKN2A $\leftrightarrow$ CDKN1A (4)}  & & - \\  \hline
MAPK1 & MAPK3 & &  &  &  & 2 \\ \hline
PDGFA & PDGFB &  &  &  &  & 2 \\ \hline
EGF & TGFA &  &  &  &  & 1 \\ \hline
\end{tabular}
\caption{Orbit equivalence classes for the glioma group.  The nodes MDM2 and TP53 can only be permuted simultaneously with the nodes CDKN1A and CDKN2A.}\label{glioma}
\end{table}

\begin{table}[h!]
\begin{tabular}{|l|l|l|l||c|} \hline
\multicolumn{4}{|c||}{Nodes} & Degree \\ \hline
PIK3CA & PIK3CB & PIK3CD & PIK3CG & 7\\ \hline
PIK3R5 & PIK3R1 & PIK3R2 & PIK3R3 &  5\\ \hline
PLCG1 & PLCG2 & TGFA & EGF & 2 \\ \hline
AKT1 & AKT2 & AKT3 &   & 12 \\ \hline
ARAF & RAF1 & BRAF &   & 8 \\ \hline
PRKCA & PRKCB & PRKCG &  & 3 \\ \hline
CCND1 & CDK4 & CDK6 &  &3 \\ \hline
BAD & CASP9 & FOXO3 &  & 3 \\ \hline
E2F1 & E2F2 & E2F3 &   &  1 \\ \hline
EGFR & ERBB2 &  &  & 9 \\ \hline
NRAS & HRAS & &  & 5 \\ \hline
MAP2K1 & MAP2K2 &  &  & 4 \\ \hline
SOS1 & SOS2 &  &  &   4 \\ \hline
MAPK1 & MAPK3 & &  &3 \\ \hline
STK4 & RASSF5 &  &  &  2 \\ \hline
CDK4 & CDK6 &  &  &  2 \\ \hline
\end{tabular}
\caption{Orbit equivalence classes for the NSCL group.}\label{NSCL}
\end{table}

\begin{table}[h!]
\begin{tabular}{|l|l|l|l|l|l|l|l||c|} \hline
\multicolumn{8}{|c||}{Nodes} & Degree \\ \hline
PIK3R5 & PIK3CA & PIK3CB & PIK3CD & PIK3CG & PIK3R1 & PIK3R2 & PIK3R3 & 3 \\ \hline
FIGF & PGF & VEGFA & VEGFB & VEGFC   &  &  &  & 4 \\ \hline
MAPK1 & MAPK3 & MAPK8 & MAPK9 & MAPK10 &  &  &  & 1 \\ \hline
AKT1 & AKT2 & AKT3 &  &   &  &  &  & 8 \\ \hline
CHUK & IKBKB & IKBKG &  &   &  &  &  & 5 \\ \hline
RAC1 & RAC2 & RAC3 &  &   &  &  &  & 4 \\ \hline
ARAF & RAF1 & BRAF &   &  &  & &  & 2 \\ \hline
TGFB1 & TGFB2 & TGFB3  &  & & &  &  & 2 \\ \hline
E2F1 & E2F2 & E2F3 &  &   &  &  &  & 1 \\ \hline
RELA & NKFB1 &  & &  &  &  &  & 14 \\ \hline
STAT1 & STAT3 &  &  &   &  &  &  & 6 \\ \hline
TGFBR1 & TGFBR2 &   &  & & &  &  & 6 \\ \hline
BAD & BCL2L1 &  &   &  &   &  &  & 4 \\ \hline
RALA & RALB &  &  & &  &  &  & 3 \\ \hline
CDK4 & CDK6 &  &  & &  &  &  & 3 \\ \hline
SMAD2 & SMAD3 &  &  & &  &  &  & 3 \\ \hline
BRCA2 & RAD51 &  &  & &  &  &  & 1 \\ \hline
TGFA & EGF &  &  & &  &  &  & 1 \\ \hline
\end{tabular}
\caption{Orbit equivalence classes for the pancreatic cancer group.}\label{PancC}
\end{table}

\begin{table}[h!]
\begin{tabular}{|l|l|l|l|l|l|l|l||c|} \hline
\multicolumn{8}{|c||}{Nodes} & Degree \\ \hline
ARNT	 &ARNT2&CREBBP&EP300&EGLN2&EGLN3&EGLN1&SLC2A1 &    \multirow{3}{*}{2}\\
FIGF	& PGF&VEGFA&VEGFB&VEGFC&TGFB1&TGFB2&TGFB3&  \\
PDGFB&TGFA & & & & & & & \\ \hline
PIK3R5 & PIK3CA & PIK3CB & PIK3CD & PIK3CG & PIK3R1 & PIK3R2 & PIK3R3 & 4 \\ \hline
PAK1 & PAK2 & PAK3 & PAK4 & PAK6   & PAK7 &  &  & 2 \\ \hline
AKT1 & AKT2 & AKT3 &  &   &  &  &  & 8 \\ \hline
ARAF & RAF1 & BRAF &   &  &  & &  & 2 \\ \hline
EPAS1 & HIF1A &  &  & &  &  &  & 18 \\ \hline
RAC1 & CDC42 &   &  & & &  &  & 6 \\ \hline
MAPK1 & MAPK3 & & & &  &  &  & 5 \\ \hline
MAP2K1 & MAP2K2 & & & &  &  &  & 5 \\ \hline
ETS1 & JUN &  &  & & &  &  & 2 \\ \hline
CRK & CRKL & &  &  &  &  &  & 2 \\ \hline
SOS1 & SOS2 &  &  &   &  &  &  & 1 \\ \hline
RAP1A & RAP1B &  &  & &  &  &  & 1 \\ \hline
\end{tabular}
\caption{Orbit equivalence classes for the renal  group.}\label{renal}
\end{table}

\begin{table}[h!]
\begin{tabular}{|l|l|l|l|l|l|l|l||c|} \hline
\multicolumn{8}{|c||}{Nodes} & Degree \\ \hline
LAMA1&LAMA2&LAMA3&LAMA4&LAMA5&LAMB1&LAMB2&LAMB3&	    \multirow{3}{*}{6}\\
LAMB4&LAMC1&LAMC2&LAMC3&COL4A1&COL4A2&COL4A4&COL4A5&  \\
COL4A6&FN1 & & & & & & & \\ \hline
TRAF1&TRAF2&TRAF3&TRAF4&TRAF5&TRAF6	& & &   \multirow{3}{*}[0.5em]{2}\\
PTGS2&NOS2 & BIRC2&BIRC3&XIAP&BCL2L1& &   & \\ \hline
PIK3R5 & PIK3CA & PIK3CB & PIK3CD & PIK3CG & PIK3R1 & PIK3R2 & PIK3R3 & 4 \\ \hline
PIAS1 & PIAS2 & PIAS3 & PIAS4 &  &  & &  & 3 \\ \hline
AKT1 & AKT2 & AKT3 &  &   &  &  &  & 11 \\ \hline
CHUK & IKBKB & IKBKG &  &   &  &  &  & 4 \\ \hline
CCNE1 & CCNE2 & CDK2 &  &   &  &  &  & 2 \\ \hline
E2F1 & E2F2 & E2F3 &  &   &  &  &  & 1 \\ \hline
RELA & NKFB1 &  & &  &  &  &  & 13 \\ \hline
MAX & MYC &  &  &   &  &  &  & 8 \\ \hline
CDK4 & CDK6 &  &  &   &  &  &  & 4 \\ \hline
APAF1 & CYCS &  & &  &  &  &  & 1 \\ \hline
\end{tabular}
\caption{Orbit equivalence classes for the SCL  group.}\label{SCL}
\end{table}

\begin{table}[h!]
\begin{tabular}{|l|l|l|l|l|l|l|l||c|} \hline
\multicolumn{7}{|c||}{Nodes} & Degree \\ \hline
RET&CCDC6&NCOA4&TFG&NTRK1&TPM3&TPR & 3 \\ \hline
LEF1 & TCF7 & TCF7L1 &  TCF7L2 &   &  &  &  3 \\ \hline
KRAS & NRAS & HRAS &   &  &   &  &   8 \\ \hline
RXRA & RXRB & RXRG &  &   &  &  &  2 \\ \hline
MYC & CCND1 &  & &  &  &  & 4 \\ \hline
MAP2K1 & MAP2K2 &  & &  &  &  &  3 \\ \hline
PPARG & PAX8 &  & &  &  &  &  3 \\ \hline
MAPK1 & MAPK3 & & & &  &  & 2 \\ \hline
\end{tabular}
\caption{Orbit equivalence classes for the thyroid cancer group.}\label{thyroid}
\end{table}

\end{document}